\documentclass[iop,apj]{emulateapj} 

\usepackage{epsfig}
\usepackage{amsmath}
\usepackage{amssymb}
\usepackage{natbib}
\usepackage{threeparttable} 
\usepackage{booktabs}

\usepackage{epstopdf}

\newcommand{\swift}{\textit{Swift}}
\newcommand{\rxte}{\textit{RXTE}}
\newcommand{\xmm}{\textit{XMM-Newton}}
\newcommand{\inte}{\textit{INTEGRAL}}

\newcommand{\Msun}{\mathrm{M}_{\odot}}
\newcommand{\lum}{\mathrm{erg~s}^{-1}}
\newcommand{\flux}{\mathrm{erg~cm}^{-2}~\mathrm{s}^{-1}}

\newcommand{\cnts}{\mathrm{counts~s}^{-1}}
\newcommand{\mdot}{\mathrm{M_{\odot}~yr}^{-1}}
\newcommand{\nh}{\mathrm{cm}^{-2}}

\newcommand{\source}{Swift J1749.4--2807}
\newcommand{\exo}{EXO 1745--248}

\slugcomment{Received 2012 April 27; accepted 2012 July 13}
\shorttitle{\source\ in quiescence}
\shortauthors{Degenaar, Patruno \& Wijnands}

\begin{document}

\title{The quiescent X-ray properties of the accreting millisecond X-ray pulsar and eclipsing binary Swift J1749.4--2807}

\author{N. Degenaar$^{1,}$\altaffilmark{3}, A. Patruno$^{2}$, and R. Wijnands$^{2}$}
\affil{$^1$Department of Astronomy, University of Michigan, 500 Church Street, Ann Arbor, MI 48109, USA; degenaar@umich.edu\\
$^2$Astronomical Institute "Anton Pannekoek," University of Amsterdam, Postbus 94249, 1090 GE Amsterdam, The Netherlands\\}

\altaffiltext{3}{Hubble fellow}

%%%%%%%%%%%%%%%
% ABSTRACT
%%%%%%%%%%%%%%%

\begin{abstract}
\source\ is a transient neutron star low-mass X-ray binary that contains an accreting millisecond X-ray pulsar spinning at 518 Hz. It is the first of its kind that displays X-ray eclipses, which holds significant promise to precisely constrain the mass of the neutron star. We report on a $\simeq$105 ks long \xmm\ observation performed when \source\ was in quiescence. We detect the source at a 0.5--10 keV luminosity of $\simeq$$1 \times 10^{33}~(D/\mathrm{6.7~kpc})^2~\lum$. The X-ray light curve displays three eclipses that are consistent in orbital phase and duration with the ephemeris derived during outburst. Unlike most quiescent neutron stars, the X-ray spectrum can be adequately described with a simple power law, while a pure-hydrogen atmosphere model does not fit the data. We place an upper limit on the 0.01--100 keV thermal luminosity of the cooling neutron star of $\lesssim$$2\times10^{33}~\lum$ and constrain its temperature to be $\lesssim$$0.1$~keV (for an observer at infinity). Timing analysis does not reveal evidence for X-ray pulsations near the known spin frequency of the neutron star or its first overtone with a fractional rms of $\lesssim$$34\%$ and $\lesssim$$28\%$, respectively. We discuss the implications of our findings for dynamical mass measurements, the thermal state of the neutron star and the origin of the quiescent X-ray emission. \\
\end{abstract}


%%%%%%%%%%%%%%%
% INTRODUCTION
%%%%%%%%%%%%%%%

\section{Introduction}
Neutron stars in low-mass X-ray binaries (LMXBs) accrete matter from a sub-solar ($\lesssim$$1~\mathrm{M_{\odot}}$) companion star that overflows its Roche-lobe. Such binaries are generally thought to be very old ($\sim$$10^{9}$~yr) and their neutron stars may have gained substantial mass during their evolution via accretion. Studying LMXBs therefore holds great promise to probe the high end of the mass distribution of neutron stars, which provides strong constraints on the equation of state of ultra-dense matter \citep[][]{lattimer2001}.

Many LMXBs are \textit{transient} and spend the majority of their time in a dim, quiescent state at a 0.5--10 keV luminosity of $L_{\mathrm{X}}\sim10^{31-33}~\lum$ \citep[e.g.,][]{jonker2004}. They exhibit occasional X-ray outbursts during which their intensity rises to $L_{\mathrm{X}}\sim10^{36-38}~\lum$, and that typically last a few weeks or months \citep[e.g.,][]{chen97}. The outbursts result from a sudden strong increase in the mass-accretion rate onto the neutron star, whereas little or no accretion takes place in quiescence. 

Accreting millisecond X-ray pulsars (AMXPs) form a small (counting 14 members to date) subclass of transient LMXBs that display coherent X-ray pulsations with a frequency of 182--599 Hz \citep[][]{wijnands2006_amxp_review,patruno2010}. Their magnetic field of $B\sim10^{8-9}$~G is strong enough to concentrate the accretion flow onto the magnetic poles of the neutron star, which creates local hotspots and gives rise to the distinctive X-ray pulsations.\\

%% Swift J1749.4--2807
\subsection{\source}\label{subsec:source}
\source\ was discovered with \swift\ on 2006 June 2 when it exhibited a thermonuclear X-ray burst \citep[][and references therein]{wijnands09}. The observed burst peak put the source at a distance of $D \lesssim 6.7$~kpc. Its 0.5--10 keV luminosity rapidly declined from $L_{\mathrm{X}}\sim10^{36}$ to $\sim$$10^{33}~(D/6.7~\mathrm{kpc})^2~\lum$ within 1 day after its discovery \citep[][]{beardmore2006,wijnands09}. The source was serendipitously detected in quiescence at $L_{\mathrm{X}}\sim10^{33}~(D/6.7~\mathrm{kpc})^2~\lum$ during three archival \xmm\ observations performed in 2000 and 2006 \citep[][]{halpern2006}. However, it was located far off-axis, inhibiting an accurate determination of the X-ray flux and spectrum \citep[][]{wijnands09}.

On 2010 April 10, renewed activity was detected from \source\ with \inte\ \citep[][]{pavan2010}. The outburst lasted for $\simeq11$~days and had an average 0.5--10 keV luminosity of $L_{\mathrm{X}}\sim10^{36}~(D/6.7~\mathrm{kpc})^2~\lum$ \citep[][]{altamirano2011_amxp,ferrigno2011}. The discovery of 518 Hz coherent X-ray pulsations with \rxte\ identified the source as a new AMXP \citep[][]{altamirano2010_amxp}. It is the first of its class to display X-ray eclipses, which occur when the companion star moves into our line of sight and obscures the central X-ray source. The eclipses recur at the $8.82$-hr orbital period of the binary \citep[][]{markwardt2010_atel,markwardt2010}.

The mass donor in \source\ is a main sequence star of spectral type K--G  \citep[][]{markwardt2010,avanzo2011}. This holds good prospects for studying the radial velocity curve of the companion in quiescence. The detection of X-ray eclipses tightly constrains the binary inclination to $i\simeq74^{\circ}-78^{\circ}$, which provides a key ingredient for such studies \citep[][]{markwardt2010}. 

Investigation of the quiescent X-ray properties is an important aspect of the challenge to accurately constrain the neutron star mass. The quiescent luminosity provides the diagnostic to determine if the donor star suffers from X-ray irradiation, which would affect its radial velocity curve \citep[e.g.,][]{bassa09}. Furthermore, the X-ray spectrum of quiescent LMXBs can be used to constrain the neutron star temperature, which offers an alternative way to probe the equation of state \citep[e.g.,][]{wijnands2001,yakovlev03,page2004,heinke2009}. In this paper we report on a long pointed \xmm\ observation of \source\ in quiescence.

 \begin{figure}[t]
 \begin{center}
	\includegraphics[width=8.0cm]{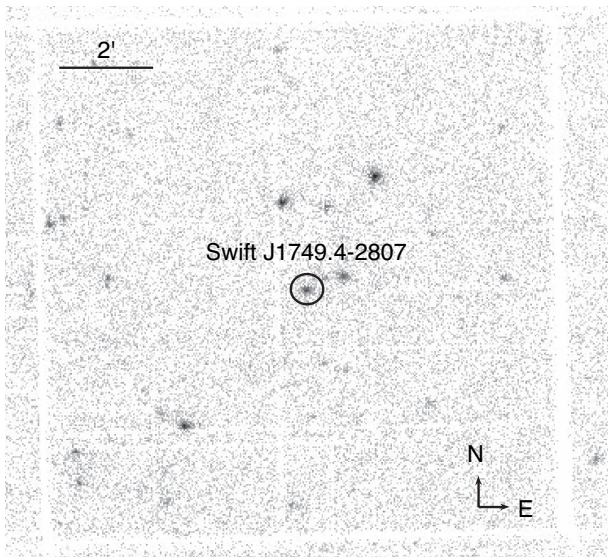}
    \end{center}
    \caption[]{Combined MOS1/MOS2 X-ray image (0.3--12 keV) of the field around \source.}
 \label{fig:ds9}
\end{figure}

%%%%%%%%%%%%%%%%%%
% OBSERVATIONS + RESULTS
%%%%%%%%%%%%%%%%%%

\section{Observations, data reduction and results}
\source\ was observed with \xmm\ for $\simeq$105~ks from UT 2011 March 19 16:25 till March 20 21:30 (Obs ID 0655670101). We used the data obtained with the European Photon Imaging Camera (EPIC). This instrument consists of two MOS detectors \citep[each made up of an array of 7 CCDs;][]{turner2001_mos}, and one PN camera \citep[an array of 12 CCDs;][]{struder2001_pn}.

Both MOS cameras were operated in the full window imaging mode. The PN was set in timing mode, in which all the two-dimensional spatial information is collapsed into a single dimension, providing a time resolution of $30~\mathrm{\mu}$s. Data reduction and analysis was carried out using the Science Analysis Software (\textsc{sas}; ver. 11.0.0).

Part of the data was affected by background flares, particularly near the end of the observation. We excluded such episodes by selecting only data with high-energy count rates of $<$0.35$~\cnts$ for the MOS and $<$0.20$~\cnts$ for the PN. This resulted in a good exposure time of $\simeq$$80$ and $\simeq$$74$~ks for the two MOS and the PN, respectively.

 \begin{figure}[t]
 \begin{center}
	\includegraphics[width=8.0cm]{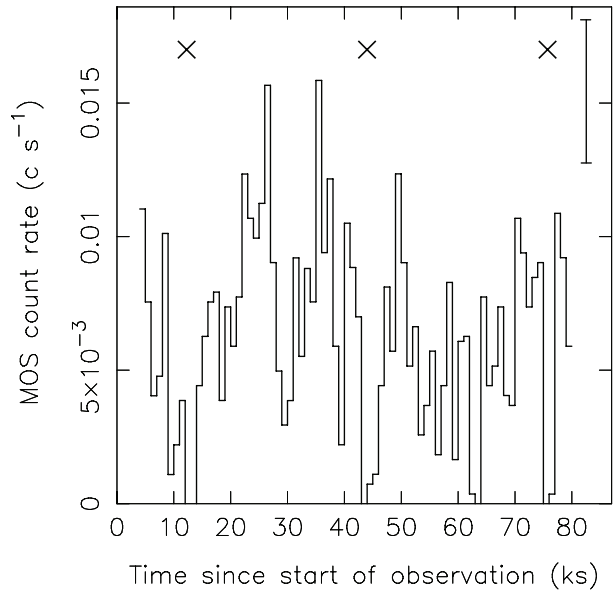}
    \end{center}
    \caption[]{Summed MOS1/MOS2 X-ray light curve at 1000 s resolution (2--8 keV). The expected times of eclipses are marked by crosses. A representative error bar ($90\%$ confidence) is shown in the upper right corner.}
 \label{fig:lc}
\end{figure}

%% IMAGE + light curve
\subsection{X-Ray Image and Light Curve}\label{subsec:lc}
Figure~\ref{fig:ds9} displays the combined MOS1/MOS2 image zoomed in on the central CCD. \source\ is clearly detected and among the brightest sources in the field. Using the tool {\sc edetect$\_$chain} we obtain a position of right ascension (R.A.) $17^{\mathrm{h}}49^{\mathrm{m}}31.58^{\mathrm{s}}$ and declination (Dec.) $-28^{\circ}08'04.5''$ (J2000.0). Combining the $0.7''$ statistical uncertainty of the detection routine with the estimated $1.5''$ MOS systematic error \citep[][]{watson2009}, yields a positional uncertainty of $1.7''$ (90\% confidence). The \xmm\ coordinates are offset by $\simeq$2.9--3.3$''$ from the 1.6--1.9$''$ \swift\ localization \citep[90\% confidence;][]{yang2010,avanzo2011}, but consistent within the errors.

We used a circular region with a radius of $12''$ to extract source events, and a $36''$-circular region placed on a source-free part of the CCD for the background. \source\ is detected in the MOS at a factor $\simeq$3 above the background ($\simeq$$1.4 \times10^{-3}~\cnts$) with an average net source count rate of $(3.3\pm0.3)\times10^{-3}~\cnts$ (0.3--12 keV).

We created background-corrected light curves for the MOS data using the tasks {\sc evselect} and {\sc lccorr}. To optimize the signal to noise ratio (S/N) we only selected events between 2 and 8 keV, since the energy spectrum of \source\ peaks in this range (Section~\ref{subsec:spec}). Figure~\ref{fig:lc} displays the summed MOS1/MOS2 light curve, which covers MJD 55639.684--55640.896. There are three strong drops in intensity that last $\simeq$2~ks and occur at times when the source was expected to be eclipsed based on its orbital ephemeris \citep[marked by crosses in Figure~\ref{fig:lc};][]{markwardt2010}. The observed times of the eclipses are consistent with the predicted ones within the resolution of the light curve ($500$~s). This firmly establishes that we have detected the quiescent counterpart of \source. During the eclipses, the source count rate drops $\simeq$$2\sigma$ below the mean and becomes consistent with zero, so no residual source flux is detected.

Apart from the eclipses, there are other instances at which the source intensity drops $\simeq$$1\sigma$--$2\sigma$ below the mean count rate (Figure~\ref{fig:lc}). There are, however, also several data points during which the count rate is $>$$1.5\sigma$ above the mean. This indicates that the additional intensity dips likely result from fluctuations in the data.

 \begin{figure}
 \begin{center}
	\includegraphics[width=8.5cm]{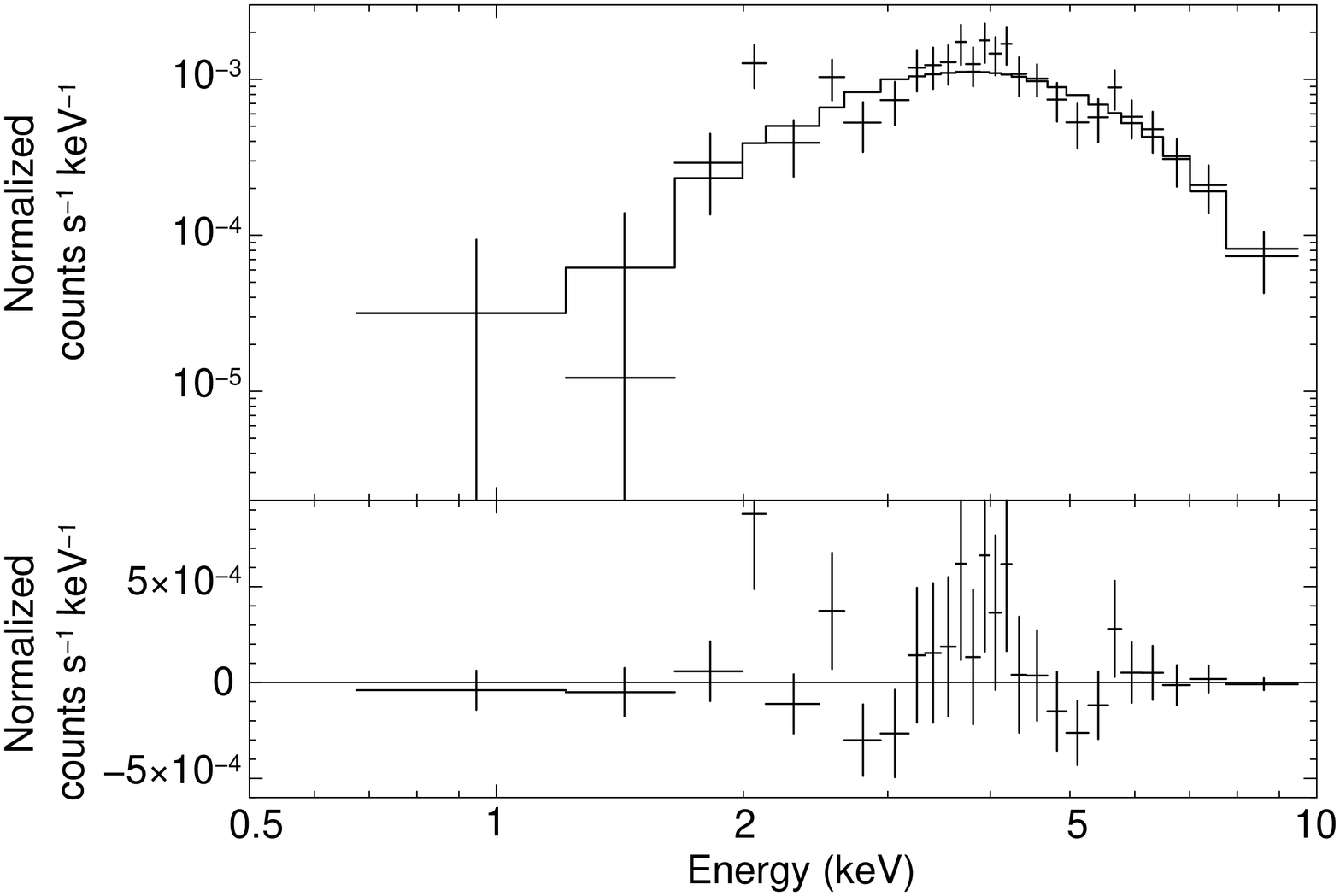}
    \end{center}
    \caption[]{Combined MOS1/MOS2 X-ray spectrum (top) and fit residuals (bottom). The solid curve represents the best fit to a single power-law model.}
 \label{fig:spec}
\end{figure}

%% SPECTRUM
\subsection{Spectral Analysis}\label{subsec:spec}
We extracted X-ray spectra from the MOS data using the task {\sc especget}, using the same source and background regions as employed for the extraction of count rates and light curves (Section~\ref{subsec:lc}). This tool also generates the appropriate ancillary response files and redistribution matrix files. We combined the two MOS spectra and the weighted response files.\footnotetext[4]{See xmm.esac.esa.int/sas/current/documentation/threads/\\epic$\_$merging.shtml.} 

The spectral data were grouped to contain a minimum of 15 photons per bin and fit in the 0.5--10 keV energy range using \textsc{XSpec} \citep[ver. 12.7.0;][]{xspec}. To calculate the correct non-eclipsed X-ray fluxes we reduced the exposure time of the averaged spectrum by 6\,500~s to take into account the presence of three eclipses \citep[][]{markwardt2010}. In all fits we include interstellar absorption along the line of sight by using the \textsc{phabs} model with the default \textsc{XSpec}-abundances and cross-sections \citep[][]{anders1989_phabs_abun,church1997}. The results of our spectral analysis are summarized in Table~\ref{tab:spec}. Quoted errors refer to 90\% confidence levels.

We first attempted to fit the spectral data with a pure-hydrogen (H) neutron star atmosphere model. We chose the model \textsc{nsatmos}, which consists of five parameters: the temperature, mass, and radius of the neutron star, the source distance, and a normalization parameter that reflects the fraction of the neutron star surface that is emitting \citep[][]{heinke2006}. We explored fits by fixing the model parameters to different values or leaving them free to float.  All trials resulted in reduced chi-squared values of $\chi_{\nu}^2 >2.4$ for 21--26 degrees of freedom (dof). We thus conclude that an H-atmosphere model does not provide an adequate description of the spectral data. 

A simple blackbody model (\textsc{bbodyrad}) does provide an acceptable fit, yielding a hydrogen column density of $N_{\mathrm{H}}=(3.6\pm1.8) \times10^{22}~\nh$, a temperature of $kT=1.6\pm0.4$~keV, and an emitting radius of $R=0.03^{+1.85}_{-0.03}$~km for an assumed distance of $D=6.7$~kpc (resulting in $\chi_{\nu}^2=1.0$ for 24 dof; Table~\ref{tab:spec}). The obtained hydrogen column density is similar to the outburst value \citep[$N_{\mathrm{H}}=3.0 \times10^{22}~\nh$;][]{yang2010,ferrigno2011}, but the best-fit temperature is considerably higher than that typically found for blackbody fits of quiescent, thermally emitting neutron stars \citep[$kT\simeq0.1-0.2$~keV; e.g.,][]{rutledge1999,wijnands2003}. This may indicate that the emission is rather non-thermal.

The data can also be fit with a single power-law model (\textsc{powerlaw}), resulting in $N_{\mathrm{H}}=(6.7\pm3.2) \times10^{22}~\nh$ and a photon index of $\Gamma=1.6\pm0.8$ ($\chi_{\nu}^2=1.0$ for 24 dof; Table~\ref{tab:spec}). This spectral shape is similar to that found for other AMXPs in quiescence (see Section~\ref{subsec:emission}). Although the obtained hydrogen column density is a factor $>$2 higher than found during outburst, the values are consistent within the errors. Figure~\ref{fig:spec} displays the combined MOS spectrum of \source\ along with the power-law fit. 

To obtain upper limits on any thermal emission due to the cooling neutron star (see Section~\ref{subsec:temp}), we added an \textsc{nsatmos} component to the best-fit power-law model. We fixed the neutron star mass and radius to $M=1.4~\Msun$ and $R=10$~km, adopt $D=6.7$~kpc, and set the normalization to 1 \citep[][]{heinke2006}, leaving the temperature as the only free fit parameter. This resulted in an upper limit on the neutron star temperature (as seen by a distant observer) of $kT\lesssim 0.1$ keV and a thermal bolometric luminosity (obtained by extrapolating the thermal model fit to the 0.01--100 keV range) of $L_{\mathrm{th,bol}}\lesssim 2 \times10^{33}~\lum$. For this fit the thermal contribution to the total unabsorbed 0.5--10 keV model flux is $\lesssim 55\%$ (90\% confidence; Table~\ref{tab:spec}).

There are hints of possible absorption features (near 3 and 5 keV; Figure~\ref{fig:spec}), which are also seen in the two individual MOS spectra. To our knowledge, there are no known instrumental features in this energy range. To estimate the significance, we add a Gaussian absorption line (\textsc{gabs}) to the broadband spectral model. This yields a central energy of $E_{\mathrm{c}}=5.1\pm0.3$~keV and a width of $0.3\pm0.3$~keV for the highest-energy feature. This implies that it is significant only at a $\simeq$2$\sigma$ level (defined as the width of the line divided by its $1\sigma$ statistical error). It has been hypothesized that gravitationally redshifted metal absorption lines may occur in the X-ray spectra of quiescent neutron stars if accretion continues at a low rate \citep[][]{brown1998}. However, such features have not been established in neutron star LMXBs with well-studied quiescent spectra.

\begin{table*}
\begin{center}
\caption{Spectral Results.\label{tab:spec}}
\begin{tabular*}{0.90\textwidth}{@{\extracolsep{\fill}}lcccc}
\hline
\hline
Parameter (unit) &  \textsc{bbodyrad} & \textsc{powerlaw} & \textsc{power law+nsatmos}  \\
\hline
$N_{\mathrm{H}}$ ($\times10^{22}~\nh$) \dotfill & $3.6\pm1.8$ & $6.7\pm3.2$ & $7.3\pm4.0$ fix  \\
$\Gamma$ \dotfill& \nodata & $1.6\pm0.8$ & $1.6\pm0.8$ \\
$kT$ (keV) \dotfill& $1.6\pm0.4$ & \nodata & $<0.1$ \\
$R$ ($D/6.7$ kpc km) \dotfill& $0.03^{+1.85}_{-0.03}$ & \nodata & 10 fix \\
$\chi_{\nu}^2$/d.o.f. \dotfill& 1.0/24 &  1.0/24 & 1.0/23 \\
$F_{\mathrm{X,abs}}$ ($\times10^{-13}~\flux$) \dotfill&  $1.0\pm0.2$ & $1.1\pm0.2$ & $1.1\pm0.1$  \\
$F_{\mathrm{X,unabs}}$ ($\times10^{-13}~\flux$) \dotfill& $1.3\pm0.1$ & $2.5\pm0.4$ & $3.2\pm0.4$ \\
$L_{\mathrm{X}}$ ($\times10^{33}~[D/6.7~\mathrm{kpc}]^2~\lum$) \dotfill& $0.7\pm0.1$ & $1.3\pm0.3$ & $1.7\pm0.2$ \\
$L_{\mathrm{th,bol}}$ ($\times10^{33}~[D/6.7~\mathrm{kpc}]^2~\lum$) \dotfill& $0.9\pm0.1$ & \nodata & $<2$ \\
Thermal fraction \dotfill& \nodata & \nodata & $<55\%$ \\
\hline
\end{tabular*}
\tablenotes{{\bf Notes.} Quoted errors represent 90\% confidence levels. The temperature is for an observer at infinity. $F_{\mathrm{X,abs}}$ and $F_{\mathrm{X,unabs}}$ represent the 0.5--10 keV absorbed and unabsorbed flux, respectively. $L_{\mathrm{q}}$ denotes the 0.5--10 keV luminosity and $L_{\mathrm{th,bol}}$ the 0.01--100 keV thermal luminosity. The bottom row gives the fractional contribution of the thermal component to the total unabsorbed 0.5--10 keV model flux. 
}
\end{center}
\end{table*}

%% TIMING
\subsection{Timing Analysis}\label{subsec:timing}
We determined the location of \source\ in the PN timing data using the tool {\sc esky2det} and accordingly extracted events using a rectangular region of $5\times199$~pixels covering the columns RAWX=37--41. A box of $5\times199$~pixels placed on RAWX=12--16 served as our background reference. We used the 2--8 keV energy range to optimize the S/N ratio (Section~\ref{subsec:spec}).

The data were folded in a single profile of eight bins by using the pulsar ephemeris reported in \citet{altamirano2011_amxp}. The ephemeris has a precision ($2\times10^{-8}$~days) that is sufficient to propagate the orbital and spin solution to the \xmm\ observation epoch ($\simeq$340 days after the ephemeris reference time). After creating the pulse profile we removed $\simeq$3\,100 background photons and were left with $\simeq$550 photons coming from the AMXP. We then fitted a sinusoid plus a constant and calculated the fractional rms amplitude of the pulsation. We repeated the folding procedure at twice the pulsar frequency to inspect the presence of a first overtone.

We did not detect pulsations with a S/N$>$3 (defined as the ratio between the pulse amplitude and its 1$\sigma$ statistical error). The 95\% confidence level upper limits are 34\% and 28\% rms for the fundamental and the first overtone, respectively. In broader energy bands the data are dominated by background noise, hence no constraining upper limits could be obtained.

%%%%%%%%%%%%%%%
% DISCUSSION
%%%%%%%%%%%%%%%

\section{Discussion}\label{sec:discussion}
We report on a long pointed \xmm\ observation of \source\ in quiescence, performed $\simeq$11 months after the end of its 2010 outburst. The source is clearly detected during our observation and we infer a 0.5--10 keV quiescent luminosity of $L_{\mathrm{X}}\simeq 1 \times 10^{33}~(D/6.7~\mathrm{kpc})^2~\lum$. This is similar to the source intensity estimated from archival \xmm\ data, which suggests that no strong quiescent variability occurred between 2000, 2006, and 2011 \citep[cf.][]{wijnands09}. The X-ray light curve shows three eclipses that are consistent in orbital phase and duration with the ephemeris derived during outburst \citep[][]{markwardt2010}. 

The quiescent X-ray spectrum of \source\ is highly absorbed ($N_{\mathrm{H}}\gtrsim 3 \times 10^{22}~\nh$) and peaks around $\simeq$4~keV. The spectral data are best fit by either a single power law with an index of $\Gamma \simeq 1.6$, or a blackbody with a temperature of $kT \simeq 1.6$~keV and an emitting radius of $R\lesssim2$~km. A pure-H atmosphere model, which usually fits the X-ray spectra of quiescent neutron star LMXBs well, does not provide an adequate description of the data. By adding an \textsc{nsatmos} component to the best-fit power-law model, we obtain an upper limit on the neutron star temperature of $kT\lesssim 0.1$~keV and on its thermal bolometric luminosity of $L_{\mathrm{th,bol}}\lesssim 2 \times 10^{33}~\lum$.  

Timing analysis did not reveal evidence for X-ray pulsations near the known spin frequency of the neutron star or its first overtone with upper limits on their rms amplitudes of $34\%$ and $28\%$, respectively. During outburst, the pulsations were detected with rms amplitudes of $\simeq$6\%--29\% (fundamental) and $\simeq$6\%--23\% (first overtone) \citep[][]{altamirano2011_amxp}. Therefore, our upper limits do not exclude that X-ray pulsations occur in quiescence.

%% X-RAY IRRADIATION
\subsection{Implications for Dynamical Mass Measurements}\label{subsec:mass}
We can asses whether X-ray irradiation may affect the radial velocity curve of the companion star in quiescence. For a binary separation of $\simeq$$2 \times 10^{11}$~cm \citep[][]{markwardt2010}, and an isotropic X-ray luminosity of $L_{\mathrm{X}}\simeq 10^{33}~\lum$ at normal incidence, we estimate an irradiation temperature at the companion star of $\simeq$2\,400~K. We consider this an upper limit, since thermal reprocessing of X-rays is likely $<$100\% efficient and the quiescent accretion disk may partly shield the companion from incident X-rays.

Isolated main-sequence stars of spectral type K--G have an effective temperature in the range of $\simeq$3\,000--6\,000~K \citep[][]{tokunaga2000}. Donor stars that have experienced mass loss and/or are rapidly rotating can, however, have a substantially lower temperature. The estimated irradiation temperature is likely too small to play an important role when compared to isolated K--G stars, but binary evolution calculations are required to asses this question robustly for the companion of \source.

%% DEEP CRUSTAL HEATING
\subsection{Thermal Emission Due to Deep Crustal Heating}\label{subsec:temp}
According to the deep crustal heating model, a neutron star is heated due to a chain of nuclear reactions that take place in the crust during outburst episodes. When accretion switches off in quiescence, the hot neutron star will thermally radiate its heat \citep[][]{brown1998}. 
Many neutron star LMXBs display soft ($\lesssim$2~keV) quiescent X-ray emission that can readily be described by a neutron star atmosphere model. The deep crustal heating mechanism is  generally accepted as the most promising interpretation of this soft thermal emission. 

The quiescent X-ray spectrum of \source\ cannot be described by a (pure H) neutron star atmosphere model. This suggests that we do not detect thermal emission from the hot, cooling neutron star. Based on the outburst history of \source\ we can estimate the thermal quiescent emission that is expected to arise from deep crustal heating, and compare this with our obtained upper limit. 

The mechanism predicts a quiescent thermal luminosity that depends on the accretion history of the binary as $L_{\mathrm{th,bol}} = \langle \dot{M} \rangle Q_{\mathrm{nuc}}/m_{\mathrm{u}} \simeq 1.9 \times 10^{18}~\langle \dot{M} \rangle~\lum$ \citep[][]{brown1998}. Here, $Q_{\mathrm{nuc}}\simeq2$~MeV is the energy released in the crust per accreted nucleon (\citealt{gupta07,haensel2008}, but see \citealt{degenaar2011_terzan5_3}), $m_{\mathrm{u}}$ is the atomic mass unit, and $\langle \dot{M} \rangle$ is the mass accretion rate onto the neutron star averaged over $\simeq$$10^4$~yr. The latter can be calculated by multiplying the average accretion-rate during outburst, $ \langle \dot{M}_{\mathrm{ob}} \rangle$, with the duty cycle of the binary (i.e., the ratio of the average outburst duration and the recurrence time).

The outbursts of \source\ appear to be short, although the exact duration is not well constrained \citep[][see also Section~\ref{subsec:source}]{wijnands09,ferrigno2011}. The average outburst intensity of $L_{\mathrm{X}}\simeq10^{36}~\lum$ suggests an accretion rate of $\langle \dot{M}_{\mathrm{ob}} \rangle = RL/GM \simeq 8 \times 10^{-11}~\mdot$ (for $R=10$~km and $M=1.4~\Msun$). An outburst length of $t_{\mathrm{ob}}=2$~weeks and a recurrence time of $t_{\mathrm{rec}}=4$~yr would imply a duty cycle of $1\%$, and a time-averaged mass-accretion rate of $\langle \dot{M} \rangle = \langle \dot{M}_{\mathrm{ob}} \rangle \times t_{\mathrm{ob}}/t_{\mathrm{rec}} \simeq 8 \times 10^{-13}~\mdot$ ($\simeq 5 \times 10^{13}~\mathrm{g~s}^{-1}$). If this is representative for the long-term accretion history of the binary, then the expected quiescent thermal luminosity is $L_{\mathrm{th,bol}}\simeq 10^{32}~\lum$. This is consistent with the upper limit obtained from our spectral analysis ($L_{\mathrm{th,bol}}\lesssim 2 \times10^{33}~\lum$; Section~\ref{subsec:spec}). Any soft thermal emission from \source\ suffers considerably from the high absorption column density. Our observational limits therefore do not place strong constraints on the thermal state of the neutron star.

We note that the low estimated long-term averaged mass-accretion rate may not reconcile with the expected mass-transfer rate from the companion found in standard binary evolution calculations \citep[e.g.,][]{podsiadlowski2002,king_wijn06}. For a Roche-lobe filling main sequence star of $M_{\mathrm{d}}\simeq0.6~\Msun$ and an orbital period of $P_{\mathrm{orb}}=8.8$~hr, the expected mass transfer rate averaged over the life time of the binary ($\simeq10^9$~yr) would be on the order of $\dot{M_{\mathrm{d}}}=\langle \dot{M} \rangle \simeq 2 \times 10^{-11}~\mdot$ \citep[see][for a review]{verbunt1993}. This is a factor $\simeq$25 higher than the mass-accretion rate that we estimated based on X-ray observations of \source. However, the mass-transfer rate from the donor star may not be equal to the mass-accretion rate onto the neutron star, e.g., because the pulsar's propeller expels matter and prevents it from accreting onto its surface \citep[][]{illarionov1975}. Furthermore, the X-ray emission may not be a good tracer of the mass-transfer rate, and the outburst behavior observed in the past decade may not be representative for the long-term accretion history of the binary. Therefore, both estimates are subject to uncertainties.

\begin{table*}
\begin{center}
\caption{Quiescent Spectral Properties of Accreting Millisecond X-Ray Pulsars.\label{tab:amxps}}
\begin{tabular*}{0.90\textwidth}{@{\extracolsep{\fill}}lccccr}
\hline
\hline
Source & $D$ & $L_{\mathrm{X}}$ & $kT$ & Thermal fraction & Reference \\ 
 & (kpc) & ($\lum$) & (keV) &  \\ 
\hline
& \multicolumn{4}{c}{Detections} & \\ 
\hline
Aql X-1$^{\dagger}$ \dotfill &  5 & $\simeq1\times10^{33}$ & $\simeq0.12$ & $\gtrsim50\%$ & 1--4 \\ 
SAX J1748.9--2021$^{\dagger}$ \dotfill & 8.5 & $\simeq 1\times10^{33}$ & $\simeq 0.09 $ & $\gtrsim60\%$ & 5,6\\ 
{\bf \source} \dotfill &  6.7 & $\simeq7\times10^{32}$ & $<0.10$ & $<55\%$ & 7 \\ 
IGR J17498--2921 \dotfill & 7.6 & $\simeq 2 \times10^{32}$ & $\lesssim0.08$ & \nodata & 8 \\ 
IGR J00291+5934 \dotfill & 4 & $\simeq1\times10^{32}$ & $\simeq0.07$ & $\simeq40\%$ & 9--12 \\  
SAX J1808.4--3658 \dotfill & 2.5  & $\simeq8\times10^{31}$ & $<0.03$ & $<5\%$ & 12--17\\  
XTE J0929--314 \dotfill & 10 & $\simeq7\times10^{31}$ & $<0.05$ & $<30\%$ & 18,19\\ 

\hline
& \multicolumn{4}{c}{Non-detections} & \\ 
\hline
IGR J17511-3057 \dotfill & 6.9 & $\lesssim 4 \times10^{33}$ & $<0.10$ & \nodata & 20 \\ 
Swift J1756.9--258 \dotfill &  8 & $\lesssim 2\times10^{33}$ & $<0.08$ & \nodata & 20,21\\  
HETE J1900.1--2455$^{\dagger}$ \dotfill &  3.6 & $\lesssim 5\times10^{32}$ & $<0.06$ & \nodata & 20,22 \\ 
XTE J1814--338 \dotfill & 8 & $\lesssim 2\times10^{32}$ & $<0.07$ & \nodata & 12 \\ 
XTE J1751--305 \dotfill & 8 & $\lesssim 2\times10^{32}$ & $<0.07$ & \nodata & 19 \\  
XTE J1807--294 \dotfill & 8 & $\lesssim 4\times10^{31}$ & $<0.05$ & \nodata & 18 \\  
NGC 6440 X-2 \dotfill & 8.5 & $\lesssim 2 \times10^{31}$ & $<0.03$ & \nodata & 20,23\\  
\hline
\end{tabular*}
\tablenotes{{\bf Notes.} The sources are arranged according to descending quiescent X-ray luminosity. The three that are marked by a dagger are ``intermittent" and do not persistently show X-ray pulsations during outburst. The quoted temperatures are for an observer at infinity and assume a pure-H neutron star atmosphere model for the thermal emission. $L_{\mathrm{X}}$ denotes the total average 0.5--10 keV quiescent luminosity (i.e., including thermal and non-thermal contributions) assuming the listed distances ($D$). The fifth column gives the fractional contribution of the thermal component to the total unabsorbed 0.5--10 keV model flux. Due to limited statistics, the spectral shape of IGR J17498--2921 cannot be constrained with the current available data.\\
{\bf References.} (1) \citealt{rutledge2001}; (2) \citealt{rutledge2002_aqlX1}; (3) \citealt{campana2003_aqlx1}; (4) \citealt{cackett2011_aqlx1}; (5) \citealt{zand2001}; (6) \citealt{cackett2005}; (7) This work; (8) \citealt{jonker2011}; (9) \citealt{jonker2005}; (10) \citealt{jonker2008_amxp}; (11) \citealt{campana2008}; (12) \citealt{heinke2009}; (13) \citealt{stella2000}; (14) \citealt{dotani2000}; (15) \citealt{wijnands2002_1808}; (16) \citealt{campana2002}; (17) \citealt{heinke2007}; (18) \citealt{campana2005_amxps}; (19) \citealt{wijnands05_amxps}; (20) \citealt{haskell2012}; (21) \citealt{patruno2010_swiftj1756}; (22) \citealt{deeg07_hetenon}; (23) \citealt{heinke2010}.
}
\end{center}
\end{table*}

%% QUIESCENT EMISSION
\subsection{The Origin of the Quiescent X-Ray Emission}\label{subsec:emission}
In addition to a soft thermal component, many neutron star LMXBs display hard emission tails that dominate the quiescent X-ray spectrum at energies above $\simeq$2--3~keV. These can typically be fit with a simple power law of index $\Gamma \simeq1-2$, which is similar to our findings for \source\ ($\Gamma \simeq 1.6$). This (non-thermal) component is not accounted for by deep crustal heating. Proposed mechanisms include residual accretion onto the magnetic field of the neutron star, while non-accretion scenarios explain the power-law emission as a shock from the pulsar wind colliding with matter flowing out of the donor star, or the pulsar wind mechanism itself \citep[e.g.,][]{campana1998,rutledge2001}.

It is interesting to compare the quiescent properties of \source\ to that of the other AMXPs. For this purpose, we have summarized the basic quiescent spectral properties of all 14 currently known sources in Table~\ref{tab:amxps} (in order of decreasing luminosity). To date, seven have been detected in quiescence and for six of these spectral analysis could be performed. Aql X-1 and SAX J1748.9--2021 are both intermittent AMXPs, which show pulsations only sporadically during their outbursts \citep[][]{altamirano2008,casella2008}. These two have the highest quiescent luminosities of $\simeq$$1 \times 10^{33}~\lum$. Although both show spectral variability in quiescence, typically $>$50\% of the emission can be ascribed to a thermal component \citep[][]{rutledge2002_aqlX1,campana2003_aqlx1,cackett2005,cackett2011_aqlx1}. 

The other AMXPs are fainter ($\simeq$$1 \times 10^{32}~\lum$ and below) and those with well-studied spectra (IGR J00291+5934, SAX J1808.4--3658, and XTE J0929--314) display power-law-dominated emission with $\lesssim$40\% attributable to a thermal component (Table~\ref{tab:amxps}). These sources also show evidence for intensity variations by a factor of a few, within single observations or between different epochs \citep[][]{wijnands05_amxps,jonker2008_amxp,heinke2009}.

\source\ is relatively bright compared to most AMXPs, but its spectral properties are not notably different: It shows a strong power-law component that can account for at least 45\% of the quiescent emission. \source\ is similarly bright as Aql X-1 and SAX J1748.9--2021. These three AMXPs have the largest orbital periods ($\gtrsim$9~hr), whereas the others have shorter orbital periods of ($\lesssim$4~hr). Unlike the other AMXPs that have been detected in quiescence on multiple occasions, \source\ does not show evidence for (strong) variations in its quiescent intensity between different epochs. However, the source was located far off-axis in the archival observations of 2000 and 2006, prohibiting an accurate determination of its quiescent flux \citep[][]{wijnands09}.

It has been noted that the quiescent spectra of the AMXPs are relatively hard compared to that of non-pulsating neutron star LMXBs \citep[][]{campana2005,campana2008,wijnands05_amxps,heinke2009}. \citet{jonker2004} studied a sample of 15 neutron stars and found a possible correlation between the fractional contribution of the non-thermal component to the total 0.5--10 keV quiescent flux. In the luminosity regime of $\simeq$$10^{33}~\lum$, the power-law contribution appears to be at its minimum and typically $<$30\%. This is lower than the $>$45\% non-thermal emission that we infer from our \xmm\ data. Although the high absorption column density strongly obscures any soft emission from \source\ and a thermal component may still contribute up to $\simeq$55\%, its spectrum indeed appears to be relatively hard compared to non-pulsating LMXBs that have similar quiescent 0.5--10 keV luminosities. 

Several other LMXBs that are similarly bright as \source\ and were observed after the study of \citet{jonker2004}, indeed also show predominantly thermal X-ray spectra with a power-law tail that contributes $\lesssim$20\% to the total quiescent emission \citep[e.g.,][]{degenaar09_exo1,servillat2012,lowell2012}. So far, the only outlier is the non-pulsating LMXB \exo\ in the globular cluster Terzan 5. This neutron star also has a quiescent luminosity of $\simeq$$10^{33}~\lum$, but displays a similarly hard quiescent spectrum as the AMXPs, with $\gtrsim$40\% attributable to a non-thermal component \citep[][]{wijnands2005,degenaar2012_1745}.
 
The fact that several of the AMXPs display particularly prominent power-law emission in quiescence, suggests a possible connection with the magnetic field of the neutron star. However, the 11-Hz X-ray pulsar in Terzan 5 is thought to have a relatively strong magnetic field ($B\simeq10^{9-10}$~G), but its quiescent emission is fully thermal and does not require the addition of a power-law component \citep[][]{degenaar2011_terzan5_2,deeg_wijn2011}. Recent observations provide strong evidence that at least in some (non-pulsating) quiescent LMXBs the non-thermal power-law emission is related to  continued accretion  that reaches the neutron star surface \citep[][]{cackett2010_cenx4,fridriksson2011}. The effect of residual accretion could be a viable explanation for the relatively luminous and hard quiescent spectrum of \source.

In addition to causing quiescent variability, residual accretion may leave detectable signatures in the X-ray emission. If the observed quiescent X-ray luminosity of \source\ can be ascribed to an accretion flow that reaches the surface of the neutron star, then the implied quiescent mass accretion rate of $ \langle \dot{M}_{\mathrm{q}} \rangle \simeq 10^{-13}~\mdot$ may allow metals to be maintained in the atmosphere \citep[][]{bildsten1992,brown1998}. This could cause the emergent spectrum to deviate from a pure H-atmosphere and give rise to gravitationally redshifted metal absorption lines. Comparing the spectral data with neutron star atmospheres of different composition has the potential to test this \citep[][]{ho2009,servillat2012}.

%%%%%%%%%%%%%%%
% ACKNOWLEDGEMENTS
%%%%%%%%%%%%%%%

\acknowledgments
N.D. is supported by NASA through Hubble Postdoctoral Fellowship grant number HST-HF-51287.01-A from the Space Telescope Science Institute, which is operated by the Association of Universities for Research in Astronomy, Incorporated, under NASA contract NAS5-26555. A.P. acknowledges support from the Netherlands Organization for Scientific research (NWO) through the Veni fellowship program, and R.W. is supported by a European Research Council (ERC) starting grant. N.D. thanks Jon M. Miller for stimulating discussions. The authors are grateful to the anonymous referee for thoughtful comments that helped improve this manuscript.

{\it Facility:} \facility{{\it XMM} (EPIC)}

%%%%%%%%%%%%%%%
% REFERENCES
%%%%%%%%%%%%%%%

\end{document}